\documentclass[11pt,a4paper]{article}

\usepackage{graphicx}
\usepackage{amssymb}
\usepackage{amsmath}
\usepackage{authblk}
\usepackage{lineno}

\begin{document}




\title{IONOSPHERIC EFFECTS DURING FIRST 2 HOURS AFTER THE CHELYABINSK METEORITE IMPACT}


\author[1]{O.I. Berngardt\thanks{berng@iszf.irk.ru}}
\author[1]{V.I. Kurkin}
\author[1]{G.A. Zherebtsov}
\author[2]{O.A. Kusonski}
\author[2]{S.A. Grigorieva}
\affil[1]{Institute of Solar-Terrestrial Physics of Sibirean Branch of Russian Academy of Sciences (ISTP SB RAS), Irkutsk, Russia}
\affil[2]{Institute of Geophysics of Ural Branch of Russian Academy of Sciences (IG UB RAS), Ekaterinburg, Russia}

\maketitle

\begin{abstract}

This paper presents the analysis of ionospheric effects in the region close to the Chelyabinsk meteorite explosion 
at 03:20UT 2013 February 15 from the Institute of Solar-Terrestrial Physics of Siberian Branch of Russian Academy of 
Sciences (ISTP SB RAS) EKB radar data, and from the Institute of Geophysics of Ural Branch of Russian Academy of Sciences 
(IG UB RAS) PARUS ionosonde data. Both instruments are located within the IG UB RAS Arti Observatory approximately 200 km 
northward from the estimated explosion site. According to the data obtained, the ionospheric disturbance caused by the 
meteorite flyby, explosion, and impact had high dynamics and amplitude. However, it obviously did not lead to a variation 
in the ionosphere mean parameters in the region above the disturbance center during the first 2 hours. Essential effects, 
however, were observed at more than 100-200 km from the explosion site and farther up to 1500 km.

\end{abstract}




\section{Introduction}

The Chelyabinsk meteorite impact at 03:20 UT on 2013 February 15, accompanied by a great number of ionospheric \cite{TertyshnikovEtAl2013,GivishviliEtAl2013}, 
atmospheric \cite{LePichonEtAl2013} and seismic \cite{TauzinEtAl2013} phenomena, will be studied for a sufficiently long time. The meteorite trajectory estimates 
are given, for example, in \cite{BorovickaEtAl2013, ZuluagaAndFerrin2013,ZuluagaEtAl2013, Proud2013}. A short review of lithospheric, magnetospheric, and atmospheric 
effects in the Asian region is presented in \cite{BerngardtEtAl2013}.

In this paper, we address the ionospheric effects accompanying the meteorite impact from the EKB radar data. The radar is an equivalent to the SuperDARN network CUTLASS 
radars \cite{LesterEtAl2004}, and is deployed at the IG UB RAS Arti Observatory ($56^{o}26'N, 58^{o}34'E$). The radar was purchased, installed, and put on a 24-h operation by 
ISTP SB RAS in mid December, 2012. This allowed one to obtain a great number of data on ionospheric conditions at the impact, and also within the period prior to and after 
the impact with a high spatial-temporal resolution. The EKB radar is located approximately 200 km northward from the explosion site. 

The radar basic operation is to concurrently observe the characteristics of the backscatter signal in the modes of oblique backscatter sounding (OBSS) and of the backscatter 
on small-scale irregularities of the ionosphere. This allows one to simultaneously estimate both the characteristics of the background ionosphere through OBSS, and the 
characteristics of small-scale irregularities through the backscatter technique.

The radar antenna system is a phased array with $~50^{o}$ scan sector, and with $~3^{o}$ lobe width. Scanning the entire sector is performed through an consecutive scan of 16 counterclockwise 
directions, approximately within 60 s, with sounding at each of the 16 fixed directions within approximately 4 s. The antenna system characteristics are such that the directional 
pattern back lobe makes is about -3dB by power of the main lobe. During the observations, the radar operated with a 60-km range resolution within the 400-3500 km range. 
The meteorite impact area is located 200 km south of the radar, in the zone of the directional pattern back lobe.

We also involved the data from vertical ionospheric sounding with the PARUS ionosonde within the Arti Observatory whose location almost coincides with the EKB radar location. 
Figure 1 shows the observational geometry.

Further in the paper, we term the bolide maximum luminosity instant (03:20:33 UT) the explosion instant, and the site where the meteorite first fragments were found (Lake Chebarkul) 
the impact site.

The meteorite impact was accompanied by effects sufficiently extended in time and space \cite{TertyshnikovEtAl2013, BerngardtEtAl2013, TauzinEtAl2013, LePichonEtAl2013, GivishviliEtAl2013, 
GorkavyyEtAl2013, EmelyanenkoEtAl2013}. However, in this paper, we address only the ionospheric effects observed during the first two hours after the meteorite impact. The region 
closest to the explosion site ($<$1500 km) is potentially the most disturbed. Therefore, studying ionospheric effects requires simultaneous high spatial-temporal resolution, which 
only an EKB radar provides currently.

\section{Ionospheric conditions}
The meteorite impact featured extremely quiet geomagnetic and seismic conditions, as well as the absence of solar flares \cite{BerngardtEtAl2013}. Except for regular disturbances 
associated with the solar terminator transit and the 2013 February 14 (previous day) effect characterized by a weak geomagnetic disturbance, the ionospheric conditions were quiet. 
Therefore, the ionospheric dynamics is to be similar to the dynamics of the quiet days near in time. This allows us to more confidently detect the effects associated with the meteorite 
flyby and impact against the regular diurnal dynamics.

To detect effects associated with regular processes in the ionosphere in the scattered signal power dynamics, we analyzed the 2013 February 15 data towards similar quiet (referential) days, 
2013 February 9-12, and 18. The Table presents the Kp-index for the days under study. From Table 1, one can see that all the selected referential days, as well as the meteorite impact day, 
were magnetically quiet days, which provided their selection.

Figure 2 exhibits the behavior of the F2-layer critical frequency (foF2) from the Arti Observatory for 2013 February 15 and for 3 referential days, 2013 February 9, 11, and 12. From these 
plots, it is apparent that foF2, estimated with the 15-min repetition period, was within the variation determined by the referential days on 2013 February 15. Thereby, the maximal electron 
density at the characteristic 15-min times within the radar vicinity during +/- 1h from the meteorite explosion differed weakly from that during the referential days. It corroborates the 
assumption of the averagely quiet ionosphere at the explosion instant. The ionospheric effects caused by the meteorite impact were observed either farther from the ionosonde location, or 
had variation characteristic times faster than 15 minutes.

\section{The 02-06 UT 2013 February 15 irregular effects}
To detect irregular effects in the EKB scattered signal power dynamics, we performed an analysis of the 2013 February 15 data towards the referential days, 2013 February 9-12, and 18. 
The main technique to detect such effects was analyzing the power of the 2013 February 15 received signal, and comparing the latter with the average power for the referential days. 
For quality estimates, we calculated the scattered signal power averaged throughout the radar entire scan sector as a function of time and range. To simplify the qualitative analysis, 
we addressed only the cases of the scattered signal high level surpassing the noise level.

Figure 3А presents a general picture of the scattered signal mean power averaged throughout the scan sector on the day of the impact (2013 February 15), whereas Figure 3B shows the mean 
power averaged by the referential days. Domains I, II where the 2013 February 15 essential deviations from the regular variation were observed are studied more scrupulously in the paper.
As seen from Figure 3A, two effects appeared most clearly defined within 02:00-06:00 UT: the effect of a powerful radial wave propagation within 04:10-05:00 UT (Effect I in Figure 3A), 
and the effect of a large-scale irregularity emergence and dynamics within 02:45-04:00 UT (Effect II in Figure 3A). The track blurring observed within 03:30-03:50UT in Figure 3B is 
associated with a sharp longitudinal dependence of the electron density and mid-scale waves around the solar terminator. The effect of a more powerful track of the signal scattered from the ground 
within 02:00-03:00 UT at 1500-2000 km does not qualitatively contradict the electron density diurnal variation. This effect is presumably associated with the previous weakly 
disturbed day (2013 February 14).

\section{The 03:45-05:00 UT radial waves in the F-layer}
Let us analyze Effect I in Figure 3А in greater detail. The analysis main technique was studying the variations in the power. The power variation was obtained by subtracting the signal mean 
level for the referential days (individually for each azimuth, time, and range) from the 2013 February 15 data. Figure 4 shows the scattered signal power for the 2013 February 15 azimuths 
0, 5, 8 and 10 (Figure 4 I-L), the mean power for 2013 February 09-12, and 18 (Figure 4 E-H), and the power variation (Figure 4A-D).

Like the analysis shows, the 2013 February 15 day was accompanied by the electron density essential disturbances having the form of oblique tracks with increasing range on the range-time 
diagram. Such peculiarities are usually interpreted as several modes of the ionospheric wave disturbances propagating at different velocities. Similar effects are observed by SuperDARN 
radars, for example, after powerful earthquakes \cite{OgawaEtAl2012}, and are associated with the transition of powerful ionospheric irregularities \cite{StockerEtAl2000}.

The qualitative analysis shows that the range to the disturbance weakly depends on the azimuth for the most powerful observed mode, and this peculiarity persists in time. This allows us 
to assume a radial propagation of the disturbance. Figure 5, for example, presents the front shape of the most powerful mode maintaining its range-time inclination from azimuth to azimuth 
and having the equivalent ionospheric velocity of ~350 m/s. Radial waves often originate in similar situations \cite{AfraimovichEtAl2001, AkhmedovAndKunitsyn2004, OgawaEtAl2012}. 
Therefore, there arise two problems: to determine the modal composition of the traveling disturbances, and to determine their center.

Because the supposed front of the irregularities is close to a spherical one, we can perform an analysis of the disturbance modal composition. To determine the disturbance modal composition, 
we summed the power variations by azimuths. Figure 6 presents the 2013 February 15 power variation averaged by all the azimuths relative to the mean power for 2013 February 9-12, and 18. 
From this Figure, one can see explicit rectilinear tracks after the explosion. We can estimate the velocities of the ionospheric disturbance different modes by these tracks.
To estimate the velocities, we build a distribution calculated by the following technique:

\begin{equation}
S(V_{0},T_{0})=\sum_{j=0..N}\sum_{i=0..N}A_{j}A_{i}\left|r_{i}-r_{j}\right|\delta(r_{i}-(r_{j}+V_{0}(t_{i}-t_{j})))\delta(T_{0}-\frac{t_{j}r_{i}-t_{i}r_{j}}{r_{i}-r_{j}})
\end{equation}

where i, j enumerate all the points on the power-range-time diagram with the power above the noise level. Each point of such a diagram features the power mean variation $A_{i}>0$ (in dB relative 
to noise), range $r_{i}$, and time $t_{i}$. Qualitatively, the formula can be explained as follows. The track is determined by the initial instant $T_{0}$, and the equivalent velocity $V_{0}$. The more the total 
area and the track power at a fixed track inclination and at the time of its start from point $r=0$, the higher the amplitude of the corresponding component in the distribution $S (V_{0}, T_{0})$. At 
numeric calculations, $\delta$-function was replaced by the equivalent rectangle whose size was determined from the necessary velocity resolution (10 m/s), and the time of the effect start (4 min).

Figure 7 shows the cut of the distribution $S(V_{0},T_{0})$ at $T_{0}$=03:20UT corresponding to the explosion time. From Figure 7, one can see that three modes with apparent velocities of 500, 800, and 
1600 m/s are robustly observed in the signal. The characteristic modes of 500 and 800 m/s have the highest amplitude, whereas the 1600 m/s mode is observed more weakly.
If we interpret the modes like the ionosphere-reflection point movement, the equivalent traveling velocity of an ionospheric point is determined by approximately half the velocity of the 
corresponding mode. This provides the estimates for radial velocities of ionospheric irregularities as 250, 400 and 800 m/s.

We can estimate the disturbance amplitude from qualitative considerations. When analyzing the tracks (Figures 4 and 6), one can see that the maximal range at the first-hop track at 04:00UT 
is approximately 400-500 km farther, than the range to the main track associated with the regular variation in the electron density. This allows us to determine the transversal size of the 
irregularity as less than 200-250 km. The electron density gain necessary to form a similar track, according to the simulation by \cite{StockerEtAl2000}, is no less than 15\%. Thereby, 
one may consider this irregularity as a meso-scale traveling ionospheric disturbance (MSTID). We will analyze the characteristics of this disturbance in a later paper.

Thus, after the meteorite impact there were a few MSTIDs with the 250, 400 and 800 m/s radial velocities in the scattered signal. The quality estimates (Figure 5) allow us to assume that 
the wave front is close to a circular one, and the center is located near the EKB radar. Figure 5 (Domain 6) presents the orientation and the direction of the irregularities.

To determine the wave center, we estimated the geometric locations of the track points for one of the main modes (400 m/s) at 04:30 UT (~70 minutes after the explosion, Figure 5 F), as 
well as the track inclination corresponding to the track geometric velocity. Table 2 summarizes the estimate results. From the estimates, one can see that the velocities are close, all 
the tracks correspond to the same mode, and the ranges at most azimuths are close to the mean value of 1290 km with a 130-km standard deviation. This means that we may consider the wave 
radial in the first approximation, and the range from the radar to the epicenter is about 130 km.

We also estimated the geometric positions of the center points by the two-point technique. The estimates allowed us to determine the approximate position of the centers shown in Figure 15 (Domain 3). 
The technique was in determining the center of the 1290-km radius (the wave mean radius from observations). This circle was constructed through two given points. From the Table 2 data set, we selected 
all the pairs and determined the circle center for each pair. Then, we defined the areas of the solutions' maximal concentration (left were only the centers the distance between which was 
no more than 10 km).

From the Figure, one can see that the suggested epicenter was south of the radar, and its location is towards the explosion site from the NASA data \cite{NASA2013}, with about a 100-km error. 
Taking into account the radar positioning precision determined by the sounding pulse characteristics and equal to 60 km, one may consider the estimate result satisfactory and being within 
the experiment error.

\section{The 02:45-04:00 UT disturbances in the E-layer}
A distinctive and unique peculiarity of the meteorite observation was the emergence of the localized ionospheric irregularity at the E-layer at 02:47 UT. The irregularity was moving east-to-west, 
had a characteristic size of about 700-800 km north-to-south and about 100-200 km west-to-east. Figure 3 shows this region as Domain II. As shown further, the existence of this irregularity allowed 
us to study the meteorite impact effects in the E-layer at small distances from the explosion site with a high temporal resolution.

Figure 3 presents the irregularity dynamics regardless of the azimuth. Figure 8 exhibits the irregularity travel dynamics in greater detail. The explicit analysis of the signal characteristics 
within 20 minutes near the explosion instant given in Figure 9 allows us to scrupulously study the dynamics of this irregularity. As seen from Figure 3, the irregularity is not regular, and it 
does not occur during quiet days. Therefore, the cause of this irregularity emergence on an exclusively quiet day, 2013 February 15, remains unclear presently. Studying the characteristics of 
this irregularity allows us to perform a detailed analysis of the processes in the E-layer at small (about 200 km) ranges from the explosion site.
This irregularity features the following parameters:

- the irregularity emerges approximately 33 minutes prior to the flare (2:47UT) southeast of the radar;

- the signal has a relatively low amplitude compared with the oblique backscatter sounding signal observed at this range;

- the irregularity travel occurs east-to-west at about 50 m/s;

- the cause of this irregularity emergence and dynamics is unknown at present.

Figure 9 shows the dynamics of this irregularity at azimuths 5-7 corresponding to the irregularity position within +/- 10 minutes from the explosion time. The axis x zero is the meteorite maximal 
luminosity moment, 3:20:33UT. We perform an analysis of the ionospheric parameters.
From Figure 9, one can see that

- the signal mean Doppler shift in the irregularity is relative low, and does not surpass 50 m/s with the prevalent direction being towards the radar (corresponds to the positive velocities), 
Figure 9 D;

- the signal spectral width is low, and below the boundary 13Hz (200 m/s) most of the time before the drop, Figure 9E;

- at the instant after the flare (first 3 minutes), one observes 2 periods of the velocity variation up to -200 m/s (Figure 9 D) with a simultaneous increase in the spectral width up to 33Hz 
(500 m/s, Figure 9E), and without signal power increase (Figure 9 C).

- after the explosion, the signal mean spectral width increased to about 13-20Hz (200-300 m/s) which evidences on a possible increase in the quantity of small-scale irregularities in the ionosphere (Figure 9 E).

In connection with a short range, one may conclude that this irregularity is obviously located at the E-layer height. The presence of relatively low velocities before the meteorite transit coinciding 
directionally with its travel direction (towards the radar), and also a comparatively small spectral width tells more likely about an oblique backscatter sounding signal, than about a signal scattered 
from the E-layer irregularities. However, a relatively low signal amplitude (compared with a well-identified OBSS signal from the F-layer) allows us to assume the signal arrival more likely from the 
back lobe, than from the main one. The back lobe attenuation is approximately -3dB, which agrees qualitatively with the relative power attenuation. Figure 15 presents the irregularity dynamics (Domain 5).

Thus, the irregularity represents a large-scale area of a higher electron density in the E-layer oriented predominately north-south, with a characteristic size in this direction not surpassing 1000 km. 
At larger size, it would also have been observed in the main lobe as a signal with a higher power. The irregularity was predominately concentrated south of the radar, and moved mostly east-to-west 
at ~50 m/s. The irregularity sizes crosswise (east-to-west) may be estimated from the power azimuthal distribution (Figure 8 С), and, presumably, did not surpass 500 km. Within this explanation, 
all the ranges observed in the experiment are to be divided approximately by two to calculate the effect location in the ionosphere. Figure 14 exhibits the effect location in the ionosphere (Domain 4).

A sharp increase in the velocity during the first minutes after the explosion may be caused by several reasons:

- by the position of the disturbance point center in the area of the observed effect;

- by the extended (not pointwise) epicenter with the characteristic size of no less than the distance from explosion site to the irregularity observation site (~200 km);

- by the disturbance high velocities allowing to travel hundreds of kilometers within a split minute (unities-tens of kilometers per second).

The most probable, from our viewpoint, is the first and the second versions because they agree well with the calculations of the center location from the radial wave transit data.

\section{Radio noise within the 10 and 16 MHz band}
Emergence of electrical phenomena is decisive for electrophonic bolides \cite{Astapovich1958, Bronshten1981}. Because the bolide, according to eyewitnesses, was electrophonic \cite{EmelyanenkoEtAl2013}, 
studying the noise background that accompanied the impact is interesting. For the EKB radar, as well as for the most SuperDARN radars, measuring the maximal background radio noise at the operational 
frequency before a sounding session is a standard procedure. This allows one to investigate the background radio noise conditions within the operational frequency in the given direction with a 60-s 
time step.

Figure 10 shows the noise amplitude depending on the time and on the azimuth. From the Figure, one can see a high temporal and azimuthal dynamics of the noise. An unusual fact is a simultaneous 
increase in the noise level at 10 and 16 MHz that occurred exactly at the explosion minute. However, we note that a similar effect during this period was also observed in other days, for example, 
on 2013 February 9 or 2013 February 11. From Figure 10, one can also see that the effect of the noise power short-term increase has, apparently, a periodic behavior at 10 MHz and 16 MHz. Therefore, 
despite the synchronism of these events, the association of this sharp increase in the noise level with the meteorite is not evident, and may be stochastic.

For a quantitative analysis, we measured the mean noise level metered with the EKB radar and scan-sector-averaged. Also, we compared the mean noise level with the data from similar measurements on 
referential days. Figure 11 shows the scan-sector-averaged noise level as a function of time in two ranges, ~10MHz and ~16MHz, measured with the EKB radar in 1-min increments.

From Figure 11, one can see that the difference of the 2013 February 15 mean noise from that of the referential days is insignificant, and practically does not surpass the triple-dispersion level 
calculated by the referential days. From the Figure 11, one can also see that there are some noise spikes surpassing these limits: 02:57UT, a simultaneous noise power decrease; 03:47UT, a 
simultaneous noise power increase. Presumably, both events are not associated with the meteorite explosion because they are substantially remote from the latter in time. Thus, one may consider 
that, at the explosion instant, the bolide did not cause any increase in the maximal noise mean level (in the radar scan sector) surpassing its day-to-day variation level within the 10 and 
16MHz bands.

\section{Ionospheric effects during the first seconds after the explosion}
We should note that at the meteorite explosion instant, 03:20 UT, there was observed a noise burst at 10 MHz that was not apparently associated with the meteorite transit which does not enable 
to tell confidently about the presence or absence of essential ionospheric effects at the explosion instant. To detect possible effects under the noise, we used special algorithms to select a 
significant signal against the referential one. Several minutes prior to and after the explosion were selected as referential. Also, we selected a few periods when the signal level at 10MHz
substantially increased.

One may consider the brightest effect accompanying the impact a sharp change in the characteristics of the received scattered signal at the explosion instant (Figure 12 C). The effect is in 
simultaneous increase in the noise and the scattered signal emergence at the ranges close to the radar within 03:20-03:21UT. The scattered signal monotonically increased its range up to the 
ranges corresponding to the F-layer. 

To detect the effect under the noise, we analyzed the power deviation from the 15-min averaged value, and subtraction from the result the background noise regular over all the ranges. The 
operation was done individually for each fixed azimuth.

Because we regard the effect comparatively large-scale, we studied the dependence of the scattered signal power as a function of the radar range and time, with no azimuth taking into account. 
With the time of obtaining correlation functions at a fixed azimuth in the experiment being about 4 s, it is the further temporal resolution for all the presented estimates.

Figure 13 exhibits the results for detecting the effect combining all the azimuths on one plot. From Figure 13, one can see that, after the emergence at 03:20:32 UT, the reflection point gains 
height for ~20-40 s traveling to the radar ranges of about 1500-1800 km (which corresponds to 700-900 km ionospheric ranges). Thus, one can see that the effect (Figure 13 C) has the dynamics 
characteristic of an OBSS signal when raising the reflection point: an increase in the range of the main scattered signal with time. Figure 14 presents the track parameters: the range to the 
signal maximal power point (Figure 14 A-E), and the maximal power proper (Figure 14 F-J). From Figure 14, one can see that the range prior to the disturbance within 03:20:32-03:21:20 UT 
increases monotonically, and the signal amplitude has a specific dynamics: it weakly varies within 3:20:32UT - 3:21:00 UT (with an ~0.3 dB dispersion), and dramatically increases by 5dB 
at 3:21:00UT. The signal smooth behavior is unusual for noise (the amplitude dispersion that is easy to determine in Figure 14 is about 1.5-2 dB). Thereby, during the remaining periods of 
the noise high level (Figure 13 A, B, D, E, F), the path dynamics differs significantly: the tracks are shorter, non-monotonic, and have a high amplitude variability.

One of possible interpretations for this effect is an assumption of forming a large-scale area with higher electron density (ionization) in the ionosphere. This area was located in the 
directional pattern back lobe, but the reasons for its formation are unknown.

Within this explanation, the signal in the directional pattern back lobe reached a range corresponding to the regular scatter from the F-layer at 03:21 UT. At the same time, there appeared 
the regular scatter from the F-layer with the 5 dB-increased power on this azimuth in the main lobe. This scatter masked the effect continuation in the back lobe. Such an amplitude increase 
agrees well with the power variation due to a double relation of the directional pattern main lobe to the back one (6 dB).

This explanation also agrees with the observation of the discussed effect of movements in the E-layer (Figure 9 D) originated unusually fast (less than for one minute) sufficiently far 
(hundreds of kilometers) from the explosion site.

Thus, a more accurate investigation into the time of the effect origin allows us to assume the time of the ionospheric effect start as 03:20:32 UT +/- 4 sec, which is close to the bolide 
maximal luminosity instant \cite{BorovickaEtAl2013}.

Unfortunately, the radar does not measure the elevation angle, which is desirable for a more correct data analysis. The lack of ionospheric instruments in this area with close spatial and 
temporal resolutions does not enable to perform a more accurate analysis. Therefore, we give possible, from our viewpoint, explanations for the effect.

One may consider a track, as a result of a small-scale irregularity radial propagation at the horizontal velocity of ~12 km/s (like this was used above when analyzing the radial wave propagation), 
or a scatter at a comparatively narrow meteor trail \cite{Bronshten1981} in the antenna directional pattern back lobe. The scatter effect at the meteoric shower accompanying the Chelyabinsk meteorite 
impact was presumably observed by the SuperDARN King Salmon radar \cite{NagatsumaEtAl2013}. However, the effect observed delay compared with the explosion instant is too great.

Another interpretation of the effect is the result of vertical upward propagation of an  arisen large-scale irregularity. In this case, the vertical velocity is to be 3.7 km/s, which significantly 
exceeds the acoustic speed appears most unlikely.

From our viewpoint, the most probable is the relaxation mechanism associated with a sharp increase in the background electron density in a large spatial area and its subsequent dynamics. After the 
forced ionization source (bolide) disappears, the electron density dynamics is determined by recombination processes, and the electron density decreases during the recombination time determined by 
the characteristic processes occurring in the ionospheric plasma. Because the lifetime determined by the recombination processes significantly grows with height, the irregularity exists longer at 
higher heights, and, consequently, the signal reflection height displaces upward. Within this model, one can estimate the characteristic height where the ionization was observed by the effect 
lifetime.

We can estimate the height based on plots of the relaxation time dependence on the height, obtained, for example, in \cite{Crain1963}. These plots allow us to estimate the approximate height of 
the irregularity as 115 km. This height agrees with the qualitative estimates for the ionization maximum height. The estimates were  obtained from the comparison of the range to the regular 
OBSS signal from the F-region (having the ~250 km height (Figure 12), and the radar range before and after the explosion (1500-1800 km) with the characteristic radar range to the investigated 
signal maximum power at the explosion instant (600-700 km). This evidences that the reflection height is approximately at 1/2-1/3 of the F-layer height, which also provides estimates of about 
80-125 km.

Thus, one may assume a higher electron density area as a source of the observed effect. This area originated south of the radar at 115 km at 03:20:32 UT, had transversal size about several 
hundreds of kilometers, and disappeared for ~20-40 s due to recombination. The existence of large-scale disturbances with horizontal size of about several tens of kilometers at of large 
meteorite impact is substantiated, for example, in \cite{ShuvalovEtAl2013}.  The absence of the effects associated with ionization of the lower layers may be a consequence of the sounding 
radio wave strong absorption in the D-layer. The irregularity short lifetime in the D-layer (out of the radar temporal resolution) also complicates the EKB observing the effects of short-living 
ionization below the E-layer. 

Within this model, one can estimate the characteristics of an ionized area with good quality. From the range of the ionospheric echo determined as half of the OBSS signal envelope delay (300 km), 
and from the supposed reflection height (115 km), it is possible to minorize the electron density according to the reflection requirements at oblique drop of a test wave (10MHz). The lower 
estimate for the plasma frequency in an ionized area at 115 km is about 3MHz.

\section{Conclusion}
In this paper, we analyzed the ionospheric effects within the 100-1500 km ranges from the Chelyabinsk meteorite explosion site from the ISTP SB RAS EKB radar data, and from the IG UB RAS 
PARUS ionosonde data. Both instruments are located at the Arti Observatory, approximately 200 km north of the supposed explosion location.

The ionospheric disturbance caused by the meteorite flyby, explosion, and impact had high dynamics and amplitude. However, it did not, apparently, lead to a variation in the mean (15-min) 
ionosphere parameters in the region over the disturbance center during the first 2 hours. Essential effects, however, were observed at more than 100-200 km from the explosion site, and farther, 
up to 1500 km. Figure 15 presents the general scheme of the observed effects.

At the first minute of the explosion, 03:20-03:21 UT, a simultaneous sharp increase in the radio noise power within the radar scan sector was observed at 10 and 16 MHz. Due to the observed 
regularity of this effect, we can not explicitly associate this increase with the meteorite event. Also, the variations in the mean background radio emission within the radar scan sector at 
10 MHz and 16 MHz did not surpass the variation level for the referential days, 2013 February 9-12, and 18.

There is a high probability that the explosion and the meteorite impact were accompanied by a short-term burst in the electron density at the heights of the E-layer that recombined then. 
The disturbance in the E-layer, apparently, formed at the explosion instant in a wide spatial area in the EKB radar directional pattern back lobe (Figure 15, Domain 7). The formation of 
this ionization area corresponds to 03:20:32 UT +/- 04 s, which agrees with the data on the explosion time. Experimental data allowed us to estimate the irregularity lifetime (20-40 s), 
the characteristic sizes (500-1000 km), and calculations also allowed us to estimate the formation height (115 km), and the plasma frequency (3 MHz) of the short-living cloud with the 
higher electron density.

Within 02:47-04:00 UT, there was an unusual meso-scale irregularity in the ionospheric E-layer (Figure 14, Domain 5), absent on the referential days, 2013 February 09-12, and 18. The 
irregularity originated at 02:47 UT, 33 minutes prior to the impact, and disappeared at 04:00 UT, 40 minutes after the explosion. The drift velocity was 50 m/s east-to-west. The irregularity 
is uncharacteristic for quiet days, and the reasons for its origin and dynamics are unknown.

A sudden emergence of this irregularity allowed us to study the meteorite effects in the E-layer (Figure 14, Domain 4). Almost simultaneously with the explosion and for 3 minutes 
(03:20-03:23 UT), there was a motion away from the radar 400 km southwest of the latter (and approximately 200 km west of the explosion site) at the E-layer height with the characteristic 
velocities ~200 m/s (Figure 14, Domain 4) and high spectral width, up to 33Hz (500 m/s). A short delay of the detected effect at a significant distance from the explosion site also 
testifies to the hypothesis of a large short-living irregularity formations at the heights of the lower E-layer, with the transversal size of several hundreds of kilometers. This 
irregularity was the source of the disturbance.

The main disturbances in the F-layer were nearly radial waves with the center close to the explosion site (80-100 km south of the radar, Figure 14, Domain 6). Analyzing the experimental 
data allowed us to determine the equivalent ionospheric velocities for individual mode travel (250, 400 and 800 m/s), and to estimate the disturbance amplitude. The increase in the 
background electron density is at least 15\% with the characteristic horizontal scales of ~200 km. The first disturbance in the F-layer was observed 15 minutes after the explosion 
at 1100 km north of the radar, and it propagated away from the radar almost radially. The radial disturbances were observed up to about 80-100 minutes. The analysis allowed us to 
estimate the disturbance center position (Figure 14, Domain 3) that was 80-100 km south of the radar and close to the explosion site from the NASA data. We could also estimate the 
wave origin time as 03:20 UT.

\section{Acknowledgements}
This study was done with financial support from Project II.12.2.3 within the Basic Research Program of the Siberian Branch of the Russian Academy of Sciences, from Integration Project 
106 of the Siberian Branch of the Russian Academy of Sciences, from Project IV.12.2 within the programs by Physical Science Department at the Russian Academy of Sciences. The authors 
thank V.P. Lebedev (ISTP SB RAS) for his fruitful discussion of the paper.

\newpage
\begin{table}
\caption{3-h values for the planetary Kp-index during 2013 February 09-18}

\begin{tabular}{|c|c|c|c|c|c|c|c|c|}
\hline 
Day&
00-03UT&
03-06UT&
06-09UT&
09-12UT&
12-15UT&
15-18UT&
18-21UT&
21-24UT
\tabularnewline
\hline
\hline 
Feb 09&
2-&
2-&
1&
1-&
0&
0&
1&
0+\tabularnewline
\hline 
Feb 10&
2-&
2-&
1+&
1-&
0+&
1-&
2&
1\tabularnewline
\hline 
Feb 11&
1+&
2-&
1-&
0+&
1-&
1&
2&
2\tabularnewline
\hline 
Feb 12&
1&
0+&
0&
1-&
1+&
2-&
2-&
3-\tabularnewline
\hline 
Feb 13&
3-&
3&
1+&
2-&
1&
1+&
3+&
4+\tabularnewline
\hline 
Feb 14&
4&
4-&
3+&
3&
3-&
2&
2+&
2\tabularnewline
\hline 
Feb 15&
0&
1&
1&
1&
1-&
1+&
2&
1+\tabularnewline
\hline 
Feb 16&
1-&
0&
0&
1-&
2+&
4&
3+&
1\tabularnewline
\hline 
Feb 17&
0+&
2-&
2-&
2&
3-&
3+&
3&
2\tabularnewline
\hline 
Feb 18&
0+&
0&
0&
0+&
1-&
1-&
1&
3+
\tabularnewline
\hline

\end{tabular}

\end{table}

\newpage

\begin{table}
\caption{Estimating the ranges from the radar to the disturbance maximum at different beams by central track for 04:30UT 2013 February 15}

\begin{tabular}{|c|c|c|}
\hline 
Beam number&
Velocity (m/s)&
Range, km
\tabularnewline
\hline
\hline 
15&
-&
1300\tabularnewline
\hline
14&
350&
1500\tabularnewline
\hline

13&
375&
1500\tabularnewline
\hline

12&
-&
-\tabularnewline
\hline

11&
-&
-\tabularnewline
\hline

10&
311&
1100\tabularnewline
\hline

9&
365&
1200\tabularnewline
\hline

8&
311&
1200\tabularnewline
\hline

7&
311&
1200\tabularnewline
\hline

6&
311&
1300\tabularnewline
\hline

5&
311&
1300\tabularnewline
\hline

4&
-&
-\tabularnewline
\hline

3&
311&
1400\tabularnewline
\hline

2&
-&
-\tabularnewline
\hline

1&
-&
-\tabularnewline
\hline

0&
375&
1200\tabularnewline
\hline

\end{tabular}

\end{table}

\newpage

\begin{figure}
\label{fig1}
\includegraphics[scale=0.7]{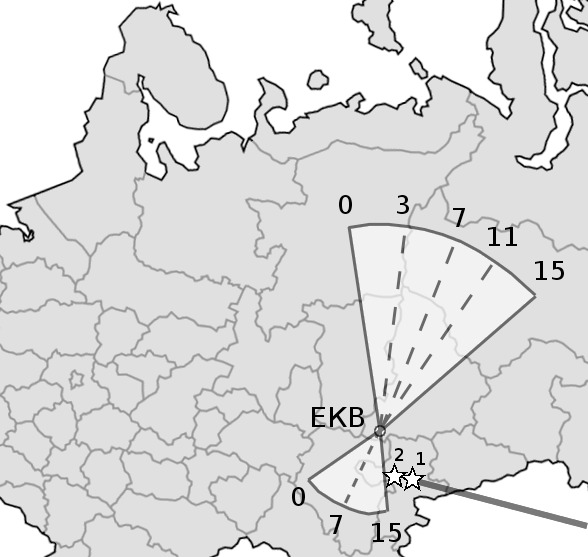}
\caption{Event geometry. A starlet 1 marks the maximal luminosity site from the NASA data. A starlet 2 marks the supposed meteorite impact site (Lake Chebarkul).
Line shows the approximate trajectory of the flyby. Circle is the EKB radar and PARUS ionosounder location. 
Sectors signify the radar scan sectors (main and back directional pattern lobes), numerals indicate beam numbers}

\end{figure}

\newpage

\begin{figure}[t]
\label{fig2}
\includegraphics[scale=0.8]{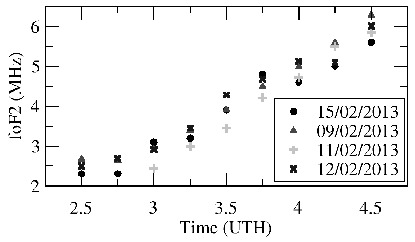}
\caption{foF2 temporal variation from the Arti Observatory data for 2013 February 09, 11, 12, 15 }
\end{figure}

\newpage

\begin{figure}[t]
\label{fig3}
\begin{center}
\includegraphics[scale=0.6]{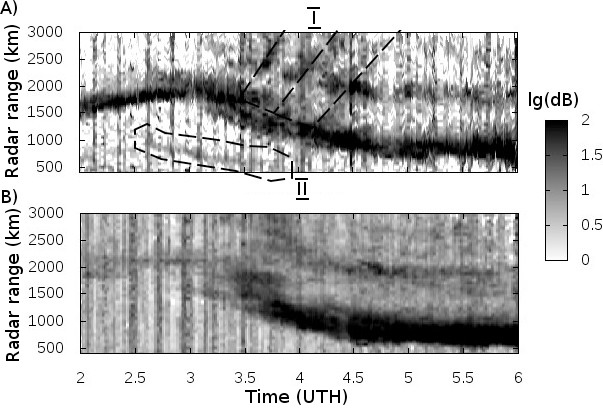}
\end{center}
\caption{
Comparison of the 2013 February 15 azimuth-averaged power (A) with the power averaged over the referential days (B). The signal-to-noise ratio is presented as log (dB)
}
\end{figure}

\newpage

\begin{figure}[t]
\label{fig4}
\begin{center}
\includegraphics[scale=0.3]{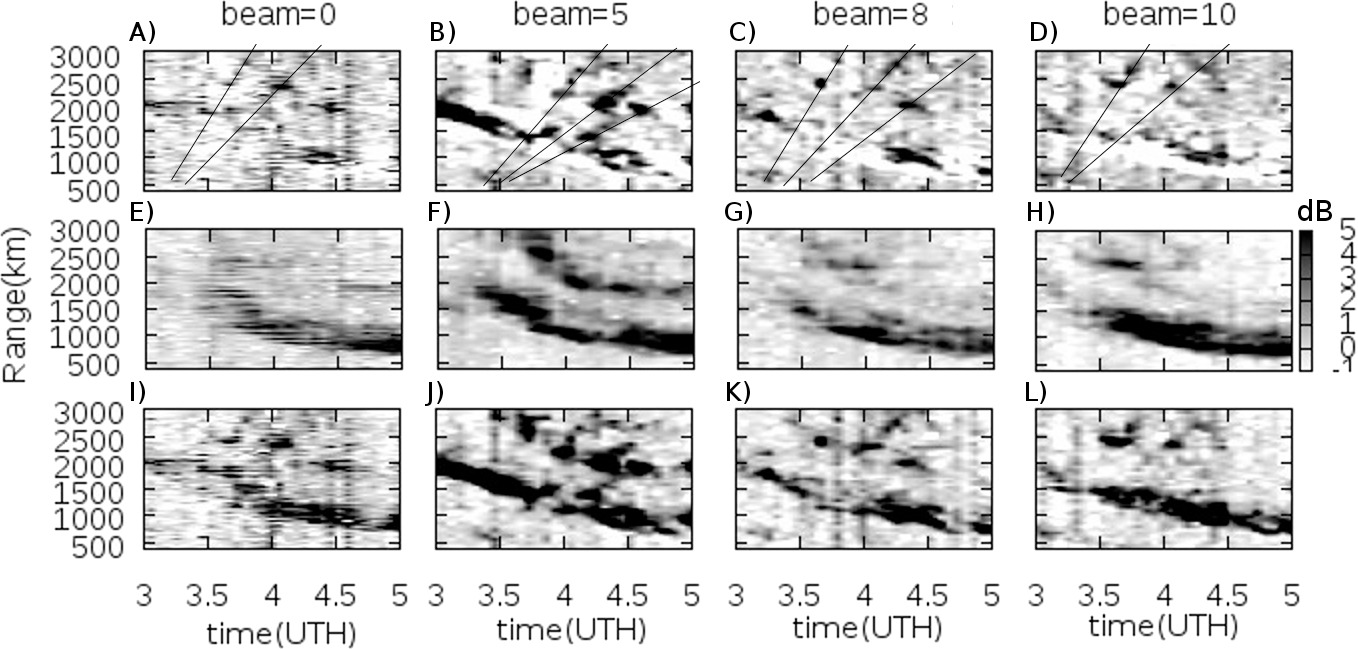}
\end{center}
\caption{
Power variation relative to the mean over the referential days, at various azimuths. The power for azimuths 0, 5, 8, and 10 on 2013 February 15 (Figure 4 I-L), the mean referential power for 2013 February 09-12, 18 (Figure 4 E-H), and the 2013 February 15 power variation (Figure 4 A-D)
}
\end{figure}

\newpage

\begin{figure}[t]
\label{fig5}
\begin{center}
\includegraphics[scale=0.5]{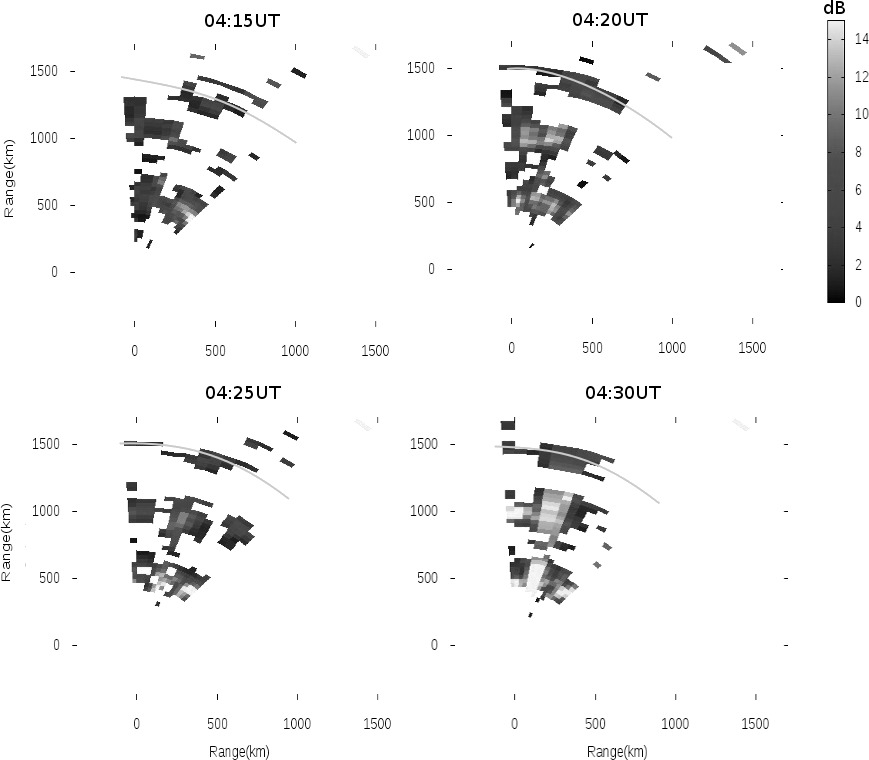}
\end{center}
\caption{
Front of one of the propagating modes having the equivalent ionospheric velocity ~400 m/s
}
\end{figure}

\newpage

\begin{figure}[t]
\label{fig6}
\begin{center}
\includegraphics[scale=0.7]{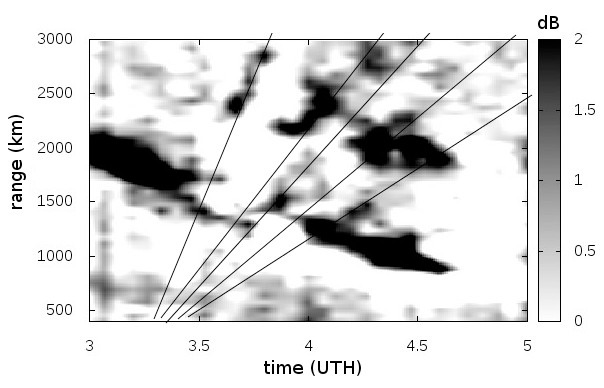}
\end{center}
\caption{
The 2013 February 15 azimuth-averaged power variation relative to the mean power for the referential days (2013 February 09-12, and 18)
}
\end{figure}

\newpage

\begin{figure}[t]
\label{fig7}
\begin{center}
\includegraphics[scale=0.7]{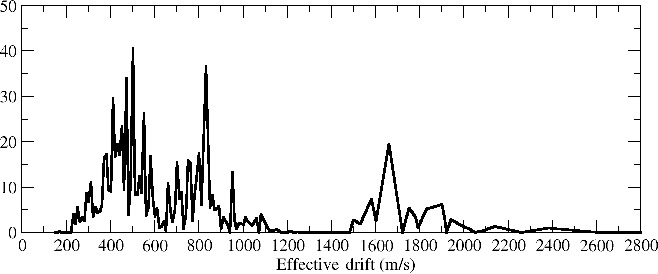}
\end{center}
\caption{
Velocity distribution $S (V_{0}, T_{0} = 03:20UT)$ for the initial instant, 03:20 UT
}
\end{figure}

\newpage

\begin{figure}[t]
\label{fig8}
\begin{center}
\includegraphics[scale=0.5]{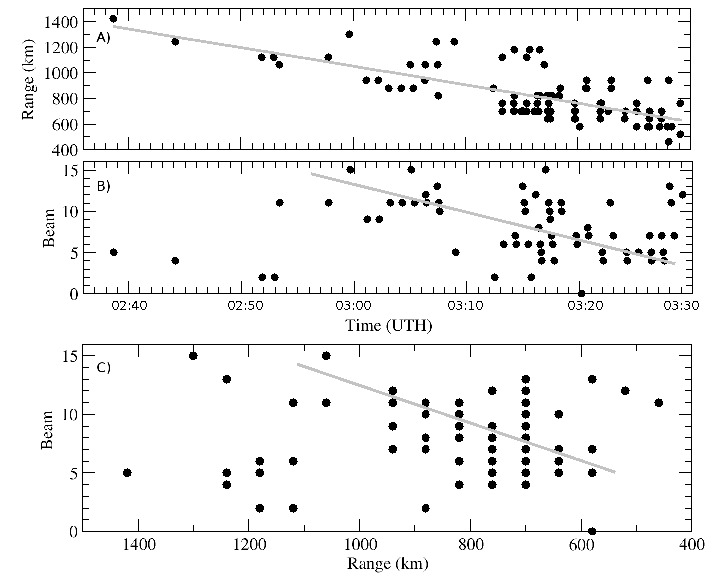}
\end{center}
\caption{
The E-layer irregularity in the directional pattern back lobe. The areas where the scattered signal arrived with the amplitude above the noise amplitude
}
\end{figure}

\newpage

\begin{figure}[t]
\label{fig9}
\begin{center}
\includegraphics[scale=0.5]{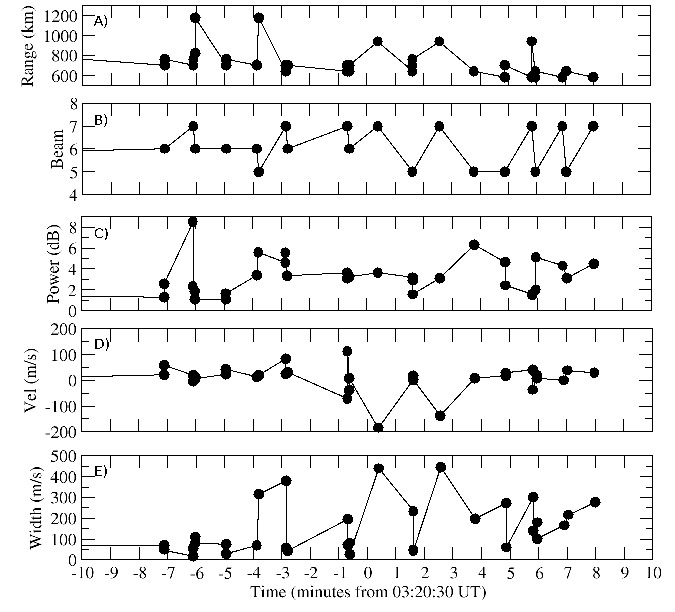}
\end{center}
\caption{
Characteristics of the signal that passed through the E-layer irregularity in the directional pattern back lobe during the meteorite impact (shown are the points corresponding to the $>$ 1 dB signal-to-noise ratio)
}
\end{figure}

\newpage

\begin{figure}[t]
\label{fig10}
\begin{center}
\includegraphics[scale=0.42]{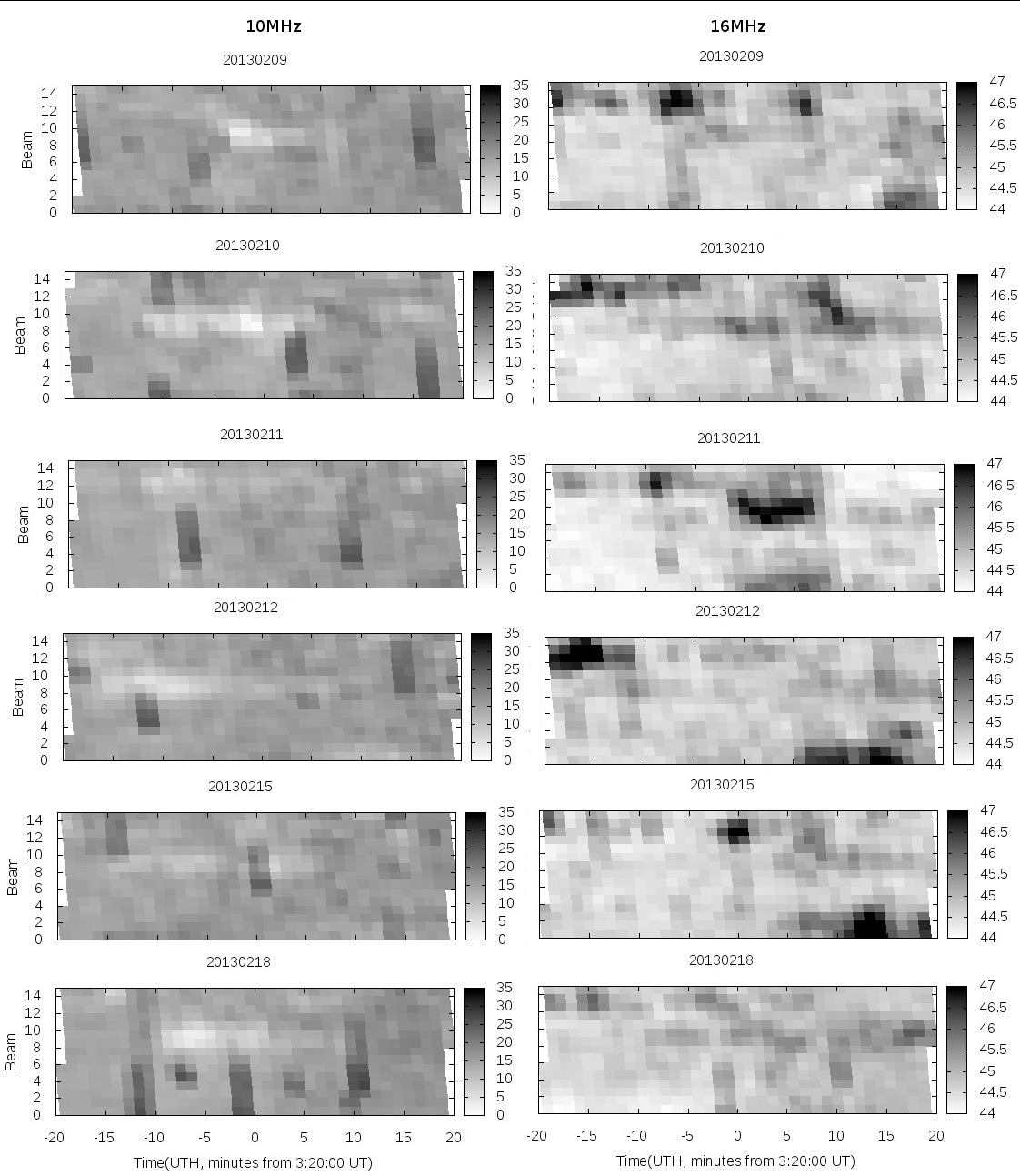}
\end{center}
\caption{
The radio noise maximal power at 10 and 16 MHz as a function of time and azimuth. Time is given in minutes relative to the 03:20UT
}
\end{figure}

\newpage

\begin{figure}[t]
\label{fig11}
\begin{center}
\includegraphics[scale=0.5]{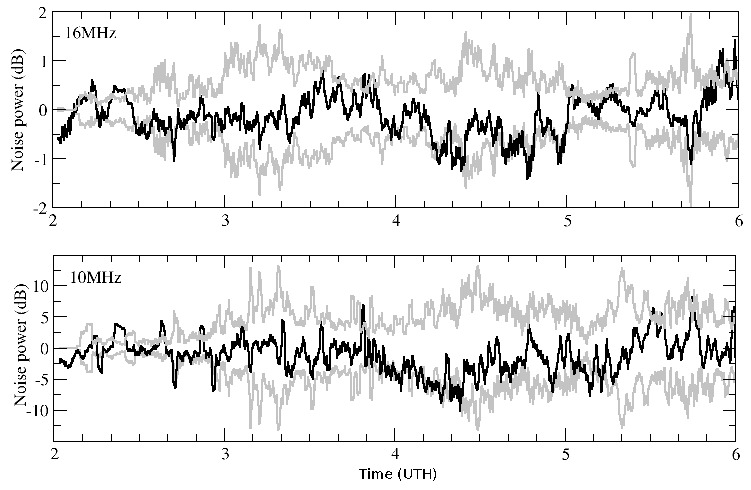}
\end{center}
\caption{
Variations in the azimuth-averaged maximal noise power at 10 and 16 MHz (black line) compared with the triple dispersion over the referential days (2013 February 09-12, and 18, grey line)
}
\end{figure}

\newpage

\begin{figure}[t]
\label{fig12}
\begin{center}
\includegraphics[scale=0.5]{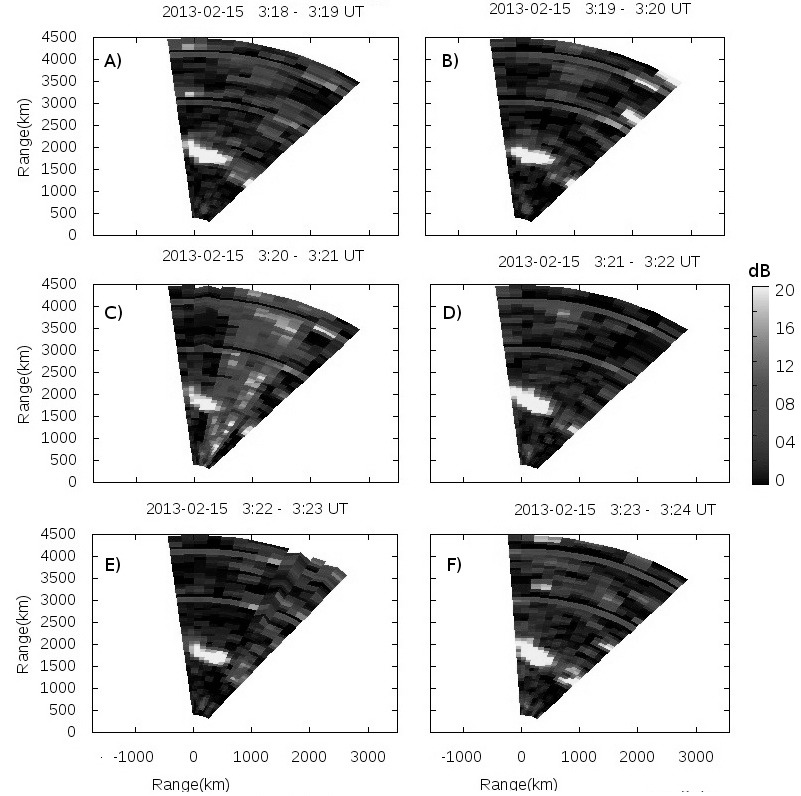}
\end{center}
\caption{
Observed short-term effect when scanning at the explosion instant (03:20 UT)
}
\end{figure}

\newpage

\begin{figure}[t]
\label{fig13}
\begin{center}
\includegraphics[scale=0.25]{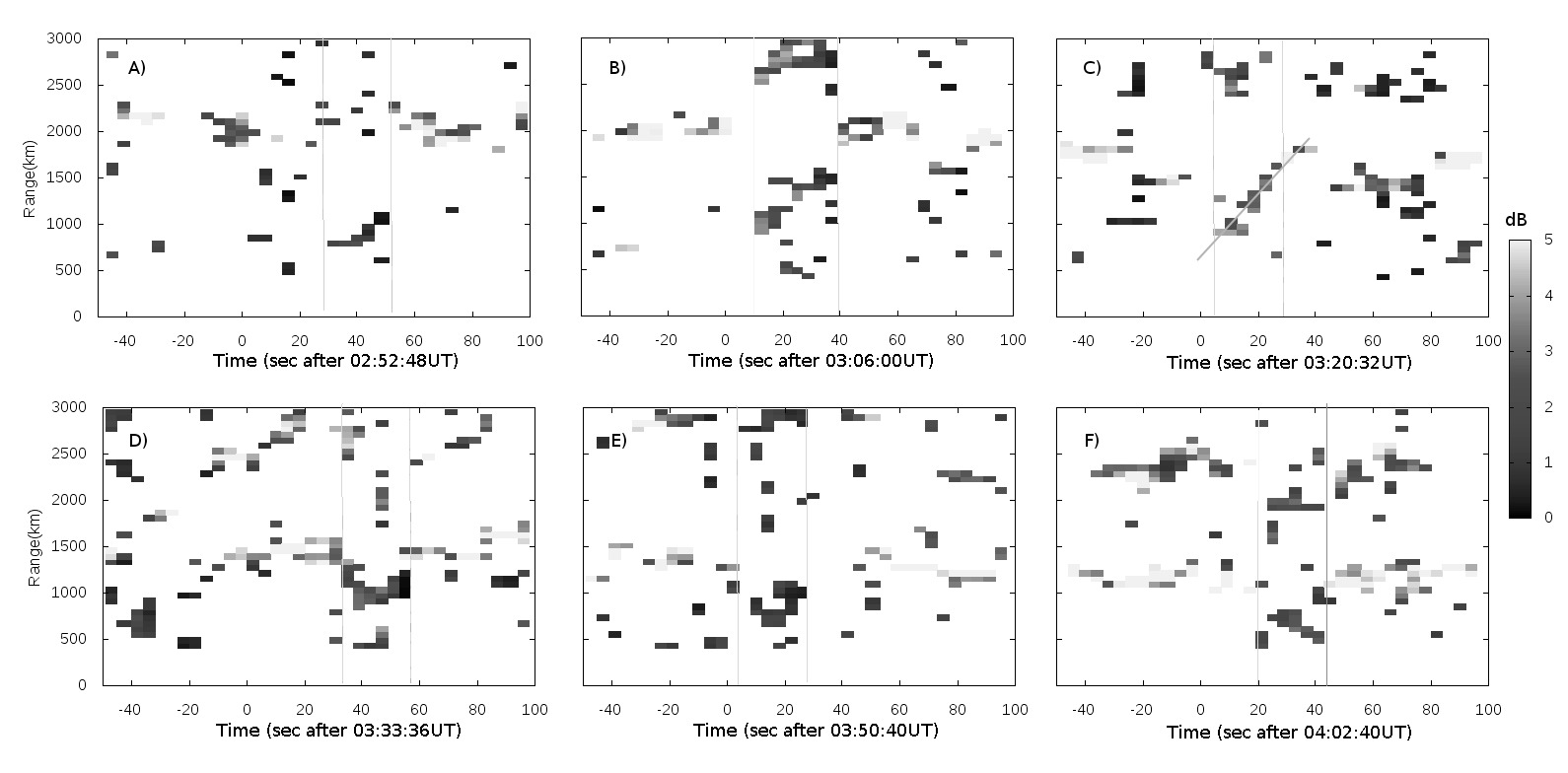}
\end{center}
\caption{
Detecting the explosion effect on the power-time-range diagram upon removing the regular component. Vertical lines mark intervals of observing the background noise level bursts at 10 MHz
}
\end{figure}

\newpage

\begin{figure}[t]
\label{fig14}
\begin{center}
\includegraphics[scale=0.7]{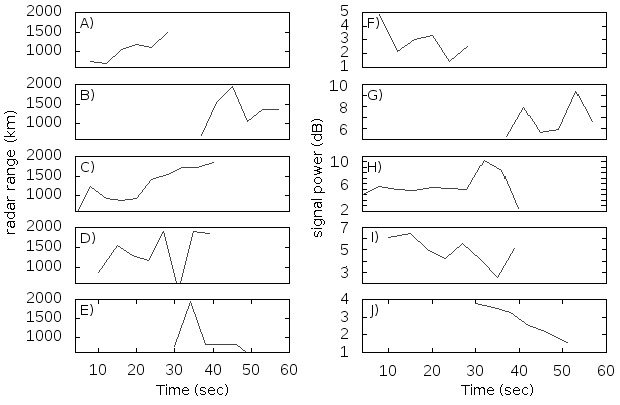}
\end{center}
\caption{
Range (A-E) and amplitude (F-J) of the signal detected against the background noise. Zero moment corresponds: 
A,F) - 02:52:48UT;
B,G) - 03:06:00UT;
C,H) - 03:20:32UT;
D,I) - 03:33:36UT;
E,J) - 04:02:40UT.
}
\end{figure}

\newpage

\begin{figure}[t]
\label{fig15}
\begin{center}
\includegraphics[scale=0.7]{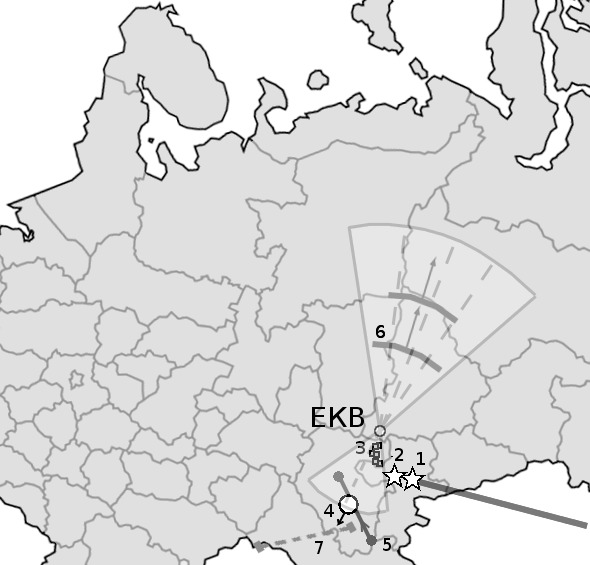}
\end{center}
\caption{
Location of the observed effects relative to the characteristic sites associated with the meteorite: 1 - the explosion site; 2 - the location of finding the first meteorite fragments; 3 - the spherical wave center from the EKB data, 03:20 UT; 4 - the wave disturbances in the E-layer, 03:23 UT; 5 - approximate trajectory of the E-layer irregularity travel, 02:47-04:00 UT; 6 - the spherical wave in the F-layer, 03:45-04:45 UT; 7 - the short-term effect of the received power increase, 03:20:32-03:21:00 UT
}
\end{figure}













\end{document}